\documentclass[10pt]{amsart} 
\begin{document} 
\title{The 
$U(1)$-invariant field theories with normal field operators} 
\author{Piotr 
\'Sniady } \address{Institute of 
Mathematics, University of Wroc\l{}aw, pl.~Grunwaldzki 2/4, 50-384 
Wroc\l{}aw, Poland \\ email: psnia\@math.uni.wroc.pl} 
\author{Marcin Zygmunt}
\address {Institute of Mathematics, Polish Academy of 
Sciences, ul.\ \'Sniadeckich 8, 00-950 Warszawa, Poland \\
email: zygmunt\@impan.gov.pl}

\begin{abstract}
In this paper we consider a very general $U(1)$-invariant field theory such 
that a field operator commutes with its adjoint, what corresponds to a 
theory of a charged bosonic particle. We show that from such an 
invariance follows the existence of particles and antiparticles associated 
to the same physical state. The field operator always turns out to 
be a sum of a particle creation and an antiparticle 
annihilation operators.  
We study in detail the case when creation and annihilation operators 
factorize and show that such operators are closely related to $q$-deformed 
commutation relations. 

\end{abstract} 

\maketitle

\newtheorem{theo}{Theorem} 
\newcommand{\Ha}{{\mathcal H}}
\newcommand{\gwia}{^{\ast}}
\newcommand{\A}{{\mathcal A}}
\newcommand{\C}{{\mathbf C}}
\newcommand{\R}{{\mathbf R}}
\newcommand{\z}{\bar{z}}
\section{ INTRODUCTION}
In quantum field theory there are today many general theorems which give 
limitations to possible field theories, e.g.\ the CPT theorem, the spin and 
statistics theorem \cite{SW}.

In this paper we investigate the implications of the 
$U(1)$-in\-va\-ria\-nce of a field theory. Our assumptions  
will be very limited, namely we consider a normal field operator $\Phi$ (an operator which commutes 
with its adjoint) such that the vacuum state is $U(1)$-invariant on the algebra  generated by
$\Phi$ and $\Phi\gwia$.
Particularly we make no assumptions 
concerning the structure of the spacetime.

We show that from such an invariance follows the existence of particles and 
antiparticles associated to the same physical state. The field operator turns 
out to be a sum of a particle creation and an antiparticle 
annihilation operators. 
We study in details the case when creation and annihilation operators 
factorize and show that such operators are closely related to $q$-deformed 
commutation relations.

Our method bases on an analysis of orthogonal polynomials in two variables \cite{Z,S}.

\section{ $U(1)$-INVARIANT FIELD THEORIES}
We consider a field theory with a Hilbert space $\Ha$ and a field operator 
$\Phi:\Ha\rightarrow\Ha$. We assume that there is a state 
$\phi:\A\rightarrow\C$ (where $\A$ is an algebra generated by operators 
$\Phi$ and $\Phi\gwia$) which is $U(1)$-invariant, i.e.\ if in any 
expression in $\A$ we replace operators $\Phi$ by $e^{i s} \Phi$ and 
$\Phi\gwia$ by $e^{-i s} \Phi\gwia$, the value of the state  should 
not change for any $s\in\R$.

Usually the state is simply a vacuum expectation: 
$\phi(S)=\langle \Omega| S\Omega\rangle$, where $\Omega$ denotes the vacuum. 
In this case the $U(1)$-invariance of a state means that the 
vacuum is $U(1)$-invariant as well, so the $U(1)$-symmetry of the theory 
remains unbroken.

Assume that $\Phi$ is a normal operator, i.e.\ 
$\Phi\Phi\gwia=\Phi\gwia\Phi$. This assumption is usually fulfilled in the 
theories of bosonic particles \cite{BS}.

By standard methods of 
functional analysis we can assign to $\Phi$ a measure $\mu$ with support on 
the complex plane 
$\C$ such that $$\phi(\Phi^k \Phi^{\ast l})=\int_\C z^k \bar{z}^l d\mu(z).$$
The $U(1)$-invariance of the state $\phi$ implies that the measure $\mu$ is 
rotation invariant.

Let us consider polynomials on the complex plane of two variables: $z$ and 
$\bar{z}$. The following theorem proven by Zygmunt \cite{Z} holds:
\begin{theo} Let polynomials $P_{k,l}(z,\z)$ form a system of orthonormal
polynomials given by Gram--Schmidt procedure applied to the sequence 
$$1,z,\bar{z},z^2,z \bar{z}, \bar{z}^2,\dots,z^n,z^{n-1} 
\bar{z},\dots,\bar{z}^n,\dots$$
with respect to the rotation invariant measure $\mu$ on $\C$ (the 
polynomial $P_{k,l}$ has the leading term proportional to $z^k \bar{z}^l$).

For all nonnegative integers $k,l$ they fulfill the following recurrence 
relations: $$z P_{k,l}=\alpha_{k,l} P_{k+1,l} + \alpha_{l-1,k} P_{k,l-1}, $$
$$\bar{z} P_{k,l}=\alpha_{l,k} P_{k,l+1} + \alpha_{k-1,l} P_{k-1,l}. $$

The coefficients $\alpha_{k,l}$ are positive for $k,l\geq 0$ and 
$\alpha_{-1,l}=0$. Furthermore for all nonnegative integers $k,l$ they 
fulfill the following relations: \begin{equation}\alpha_{k,l} 
\alpha_{l,k+1}=\alpha_{l,k} \alpha_{k,l+1}, \label{r1}
\end{equation} 
\begin{equation} 
\alpha_{k,l}^2+\alpha_{l-1,k}^2=\alpha_{l,k}^2+\alpha_{k-1,l}^2. \label{r2}
\end{equation}
 \end{theo}

Using the notation of this theorem let us denote
$$\psi_{k,l}=P_{k,l}(\Phi,\Phi\gwia)\Omega\in\Ha.$$ We have 
\begin{equation} \langle \psi_{k,l}|\psi_{m,n}\rangle=\int_\C 
\overline{P_{k,l}} P_{m,n} d\mu(z)=\delta_{k,m} \delta_{l,n}, 
\label{ortonormal} \end{equation} 
\begin{equation}\Phi 
\psi_{k,l}=\alpha_{k,l} \psi_{k+1,l} + \alpha_{l-1,k} \psi_{k,l-1}, 
\label{s2}
 \end{equation}
\begin{equation} \Phi\gwia \psi_{k,l}=\alpha_{l,k} \psi_{k,l+1} + 
\alpha_{k-1,l} \psi_{k-1,l}. \label{s3} \end{equation}

We see that $\psi_{k,l}$ is a family of orthonormal vectors in $\Ha$ indexed 
by a pair nonnegative numbers. We can think that the index $k$ represents 
the number of particles and the index $l$ the number of antiparticles which
are in the physical state corresponding to the operator $\Phi$.

Equations (\ref{s2}) and (\ref{s3}) tell us that the operator 
$\Phi=K\gwia+\Lambda $ is a sum of two operators: $K\gwia$---a creator of a 
particle and $\Lambda$---an annihilator of an antiparticle, while 
$\Phi\gwia=K+\Lambda\gwia$ is a sum of a annihilator of a particle and a 
creator of an antiparticle.

\section{ FACTORIZABLE CREATION AND ANNIHILATION
 OPERATORS}
It natural to restrict our considerations to the case when each of the 
operators $K,K\gwia,\Lambda,\Lambda\gwia$ is a product of two operators, 
each depending either on the number of particles or the number of 
antiparticles.

This means that we are solving the equation $\alpha_{k,l}=f_k g_l$. From 
the equation (\ref{r1}) it follows that 
$$g_l g_{k+1}=g_k g_{l+1}, $$ 
so $(g_i)$ form a geometric sequence. We may take it in the 
form $g_i=q^i$ since the constant factor can be moved to the $(f_i)$ 
sequence. From the assumption $\alpha_{k,l}>0$ follows that $q>0$.

Now the equation (\ref{r2}) takes the form
$$[(f_k)^2-(f_{k-1})^2] q^{-2k}=[(f_l)^2-(f_{l-1})^2] q^{-2l},$$ 
therefore there exists a constant $c$ such that 
$$[(f_k)^2-(f_{k-1})^2] q^{-2k}=c^2$$
from which follows that
$$\alpha_{k,l}=c \sqrt{1+q^2+\cdots+q^{2k}}\ q^{l}=\left\{
\begin{array}{cl} c \sqrt{\frac{1-q^{2k+2}}{1-q^2}}\ q^l & \mbox{for }  q\neq 1 \\ \\ 
c \sqrt{k+1} & \mbox{for }q=1  \end{array} \right. $$

We see that the case of $q=1$ corresponds to $\Phi$ being equal (up to a 
constant) to a sum of a classical 
bosonic particle creator and antiparticle annihilator.

In the general case we have $K=c A_k q^{N_l}$ and $\Lambda=c A_l
q^{N_k}$ where $N_k$, $N_l$ denote the number of particles or number 
antiparticles operators respectively and $A_k$, $A_l$ denote the particle 
and antiparticls annihilation operators, $A\gwia_k$, $A\gwia_l$ denote the 
particle and antiparticle creation operators. 

Operators concerning particles and 
antiparticles commute: $$A_k A_l=A_l A_k,$$
$$A_k\gwia A_l\gwia=A_l\gwia A_k\gwia,$$
$$A_k\gwia A_l=A_l A_k\gwia.$$
Creation and annihilation operators fulfill the $q^2$-deformed commutation 
relation \cite{FB}:
$$A_k A_k\gwia-q^2 A_k\gwia A_k=1$$
$$A_l A_l\gwia-q^2 A_l\gwia A_l=1$$

\end{document}